\documentclass[10pt,twocolumn,pre,byrevtex,showpacs,superscriptaddress,showkeys,bibnotes]{revtex4}
\usepackage{graphics}

\begin{document}

\noindent\parbox{\textwidth}{\small{
Proceedings of the 3rd International Conference on\\
News, Expectations and Trends in Statistical Physics (NEXT-$\Sigma\Phi$ 2005)\\
13-18 August 2005, Kolymbari, Crete, Greece}}
\bigskip

%\begin{frontmatter}

\title{Metastability in the Hamiltonian Mean Field model and Kuramoto model}

\author{ Alessandro Pluchino and Andrea Rapisarda}
\email{alessandro.pluchino@ct.infn.it, andrea.rapisarda@ct.infn.it  }
\affiliation{Dipartimento di Fisica e Astronomia and Infn Universit\'a  di Catania, Italy}

\date{\today}

\begin{abstract}

We briefly discuss the state of the art on the anomalous dynamics of the Hamiltonian Mean Field model.  We  stress the important role of the initial conditions for understanding the microscopic nature of the intriguing metastable quasi stationary states observed in the model and the connections to Tsallis statistics and glassy dynamics.  We also present new results on the  existence of  metastable states in the Kuramoto model and  discuss the similarities with those found in  the  HMF model.  The existence of metastability  seem to be quite a common  phenomenon in fully coupled systems, whose  origin could be also interpreted as a dynamical  mechanism preventing or hindering sinchronization.   

\end{abstract}

\maketitle

%\end{frontmatter}

\section{Introduction}

The Hamiltonian Mean Field (HMF) model intensively studied in the last years can be considered 
a  paradigmatic  system for understanding the behavior of long-range interacting systems
 \cite{lhs1}. In particular its striking out-of-equilibrium dynamics has raised much  
interest \cite{lrt,cmt,monte,yama,morita,gian,chava1,chava2,hmfvetri,bouchet,barre,rocha,epnhmf1,epnhmf2}
 for the anomalous dynamics connected with the existence of metastable 
quasi stationary states (QSS). Several claims have been advanced concerning a theoretical description of  this phenomenon within standard statistical mechanics \cite{yama,bouchet,barre,rocha}. However these studies neglects the important role of the initial conditions and the 
 hierarchical fractal-like structures which are generated in the $\mu$-space. In
 our opinion  the situation is still much more complex and not completely clear although some progress has certainly been made.  
In this very short paper we want summarize our point of view based on the most recent numerical simulations \cite{epnhmf1} which indicate a connection to Tsallis generalized statistics  \cite{tsa1} and glassy dynamics \cite{hmfvetri}.  We present also another interesting point of view, which could probably help in  understanding this anomalous behaviour, exploiting an analogy with recent numerical results on   metastability recently found in the  Kuramoto model. In other words, metastability could likely be seen  also  as a kind of dynamical hindrance to synchronization.

%%%%%%%%%%%%%%%%%%%%%%%%%%%%%%%%%%%%%%%%%%%%%%%%%%%%%%%%%%%%%%%%%%%%%%%%%%%%%%%%%%%%%%%%% 
\begin{figure*}[t]
\resizebox{6.0in}{!}{\includegraphics{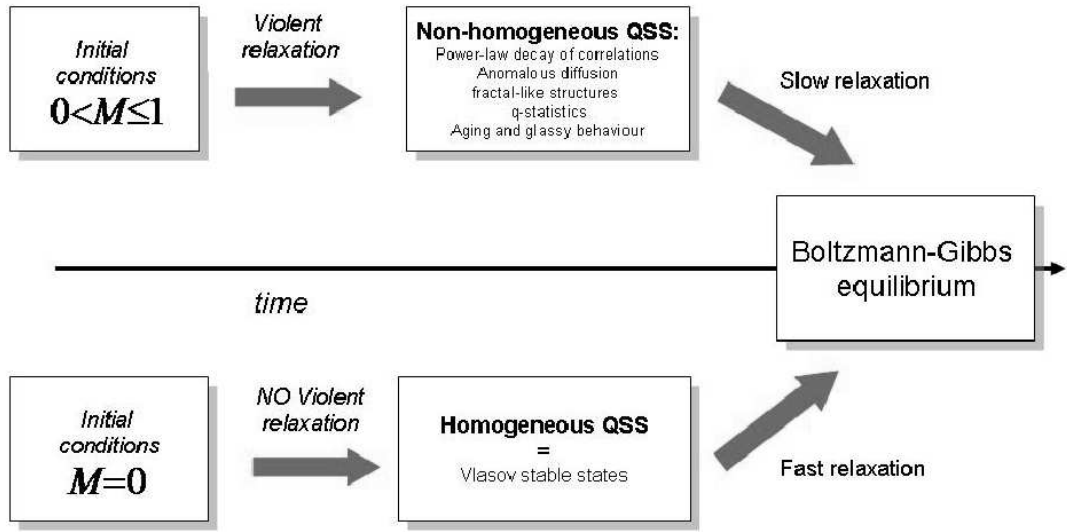}}
\caption{Schematic figure illustrating the relaxation to BG equilibrium in  the HMF model. According to the initial condition of the magnetization the nature of the QSS presents  different features and the routes to thermalization  are not the same. See text}
\label{metae1}
\end{figure*}
%%%%%%%%%%%%%%%%%%%%%%%%%%%%%%%%%%%%%%%%%%%%%%%%%%%%%%%%%%%%%%%%%%%%%%%%%%%%%%%%%%%%%%%%%%

\section{Metastability in the HMF model}
The HMF model describes a system of planar rotators or spin vectors $\overrightarrow{s}_i =(\cos \theta_i, \sin \theta_i)$
with unitary mass and interacting through a long-range potential. The Hamiltonian can be written as 
\begin{equation}
H = K + V = \sum_{i=1}^N \frac{p_i^2}{2} + \frac{1}{2N}
\sum_{i,j=1}^N [1-\cos(\theta_i - \theta_j)]~~~~,
\label{model0}
\end{equation}
being $\theta_i $ and $p_i$conjugate variables. The latter can also describe 
 a system of fully interacting particles rotating
on the unit circle. Introducing the 
{\em magnetization} 
$M = \frac{1}{N} |\sum_{i=1}^N \overrightarrow{s}_i |$ as order parameter, 
it is possible to solve the model exactly  at equilibrium, where a
second-order phase transition exists. 
The system passes from a low-energy condensed
(ferromagnetic) phase with magnetization  $M\ne0$, to a
high-energy one (paramagnetic), where spins are homogeneously
oriented on the unit circle and $M=0$. 
The critical point is found at a temperature $T_c=0.5$, which 
corresponds to the critical energy density $U_c=0.75$. The 
{\em caloric curve} is given by 
$U =H/N = \frac{T}{2} + \frac{1}{2} \left( 1 - M^2 \right)$\cite{lhs1}.
\\
At variance with this  equilibrium scenario, the out-of-equilibrium
dynamics - explored by means of numerical simulations - 
shows several anomalies in a special subcritical region
of energy values  and in particular for $0.68<U<U_c$. 
Starting from out-of-equilibrium 
initial conditions, the system  remains trapped in 
metastable quasi-stationary states with vanishing magnetization
at a temperature lower than the equilibrium one, until it slowly relaxes towards
Boltzmann-Gibbs (BG) equilibrium. This transient QSS regime becomes stable
if one takes the infinite size limit before the infinite time
limit~\cite{lrt,cmt}. In this case the system stays forever in the $M=0$ state 
at the limiting temperature $T_{QSS}$.
\\
In several previous works \cite{cmt,lrt} we have clearly 
shown that the microscopic nature of the
anomalous QSS regime depends  in a sensitive way on the choice of the initial conditions (IC).  
In this respect we focused on two main classes of IC (see Fig 1).
\\
(1) The first class of IC contains  the so-called \textit{M1 IC}, with all the angles put equal to zero, i.e.
$M(0)=1$ and a {\it water bag} (uniform) distribution of momenta.  In this class we can also consider those initial conditions  with finite magnetization, $M(0)>0$.  When the system is started with these initial conditions 
an initial {\it thermal explosion}, with a following violent relaxation, occurs.  After this stage,  the system quickly freezes and remains trapped in long-living 
QSS with vanishing magnetization and a temperature smaller than the equilibrium one. In this case  the system appears 
macroscopically quite homogeneous, since averaging over all  angles the  magnetization tends to vanish with 
the size of the system.  On the other hand {\it from a microscopic point of view it is not so}. In fact fractal-like  structures are generated in the $\mu$-space and many dynamical 
anomalies appears, among which anomalous diffusion, power-law decay of velocity correlations, aging, weak ergodicity breaking and dynamical frustration \cite{lrt,cmt}.  
Such a behavior  can be usefully described by the Tsallis' 
generalized formalism \cite{epnhmf1,tsa1} and interesting analogies with glassy systems do also exist \cite{hmfvetri}. 
\\
(2) The second main class of out-of-equilibrium initial conditions 
indicated here as  \textit{M0 IC}, see Fig.1,  is characterized by a uniform distribution of 
both angles ($M(0)=0$) and momenta (water bag). This choice puts 
the system directly in the limiting long-living QSS with zero
magnetization and temperature $T_{QSS}$.   
This is a particular case which 
corresponds to a stationary solution of the Vlasov equation \cite{yama,bouchet,barre,rocha}. 
The latter is 
stable in the thermodynamic limit but metastable for finite sizes 
of the system.  After the QSS transient, 
the system relaxes {\it very quickly } (almost exponentially) to the BG equilibrium \cite{cmt}. 
It is important to stress that although macroscopically also in 
this case we observe metastability  as in the M1IC case, 
this time the initial state is spatially very  homogeneous from the 
beginning and no violent relaxation  occurs. The 
force acting on each particle (spin) is zero since $t=0$ and all
the microscopic anomalies and correlations observed for initial conditions 
with finite magnetization are almost absent \cite{cmt}. 
\\
The very different microscopic behavior of the QSS regime 
corresponding to these two class of initial conditions strongly
suggests a different nature of the macroscopical metastability. 
In the M0IC case the Vlasov stability argument \cite{yama,bouchet} 
is rigorous  due to perfect homogeneity of the limiting 
temperature QSS. On the other hand  in the M1IC case, microscopic 
fractal-like structures created by the sudden dynamical cooling of
the system do not correspond to
 the starting homogeneity hypothesis on which the Vlasov approach used 
in \cite{yama,bouchet,barre} is based and create a much more complex situation \cite{chava1,chava2}.
In this latter case, the QSS metastability seems 
to be more related to the ergodicity-breaking phenomenon and to 
the trapping effect linked to the complex structure of the phase 
space region selected by the finite magnetization initial 
conditions. 
\\
We summarize below the main numerical results on which our analysis is based.
In refs.\cite{cmt} and elsewhere we studied the behavior of both 
velocity autocorrelation functions and anomalous diffusion
in the QSS regime for U=0.69 and for different initial conditions.
A similar slow power-law decay of velocity correlations for $ 0.4\le M(0)\le1$ 
was found, while for $M(0)=0.2$, and even more for $M(0)=0$, the decay is much  faster.  
By fitting these relaxation curves by means of the 
Tsallis' q-exponential function it was possible to characterize  in a quantitative way  
the dynamics originatined by  these  different initial conditions. 
In fact we found a value $q=1.5$
for $M(0)\ge0.4$, while we got $q=1.2$ and  $q=1.1$ for $M(0)=0.2$
and for $M(0)=0$ respectively - notice that $q=1$ corresponds
to the usual exponential decay.
On the other hand, by studying the mean square displacement
of angles  (which typically scales as $\sigma^{2}(t)\sim t^{\gamma}$), 
again for U=0.69 and as a function of the
initial conditions, it was  found superdiffusive behavior for $0.4\le M(0)\le1$
with an exponent $\gamma=1.4-1.5$. At variance,  for M(0)=0 we obtained  $\gamma= 1.2$,
a value that  tends to $\gamma\sim 1$ (normal diffusion) increasing the
size of the system.
\\
In refs.\cite{cmt} it was found a connection between 
 the velocity correlations decay and the diffusive behavior of metastable states 
 based on  the  simple relationship  $\gamma=2/(3-q)$, which 
links the   diffusion exponent $\gamma$  with  the entropic index $q$.
This formula was  derived  theoretically by Tsallis and Bukman within 
  a generalized Fokker-Planck approach
based on q-statistics \cite{tsa1}.
For various initial conditions,  ranging from
M(0)=1 to M(0)=0 and different sizes at $U=0.69$, it was found that this formula
is correct.
This latter  holds also for a generalized version of the model, the $\alpha-XY$ model, where 
the long-range character of the interaction is modulated through a term  $r^{-\alpha}$ \cite{gian}.
In such a way one can predict the value of $q$ from the knowledge of $\gamma$
and vice-versa.
\\
Another framework that seems to be very promising in order to shed light
on the microscopic nature of the QSS is that one of glassy dynamics.
The first link between QSS metastability and glassy behavior was suggested
by the discovery of the aging phenomenon in autocorrelation functions \cite{monte,cmt}
and by the {\it dynamical frustration} scenario observed in the QSS regime obtained for M1IC. 
In this  case  clusters appear and disappear on the unit circle, competing one with each other
in trapping more and more particles and generating a dynamically frustrated
situation typical of  glassy systems.
In refs.\cite{hmfvetri} it was 
 introduced a new order parameter for the HMF model, 
the {\it polarization}, in order to measure the degree of freezing of the rotators.
By means of this new parameter one can  discriminate between M1IC and M0IC metastability,
showing that only starting from the first class of initial conditions the QSS regime
could be considered (in the thermodynamic limit) as a sort of glassy-phase
for the HMF model.
More recently we introduced an effective spin-glass Hamiltonian in order to study 
this anomalous glassy-like dynamics from an analitical point of view. 
By means of the Replica formalism,  it was found a self-consistent equation 
for the glassy order parameter. The latter is  
 able to predict, in a restricted energy region below 
the phase transition, the polarization values obtained with molecular dynamics simulations \cite{hmfvetri}.
\\
All these results indicate  a deep dynamical nature of QSS metastability.
In the next section we present new results concerning  metastability in the  
Kuramoto model which shows a  certain analogy with those of the HMF model.

%%%%%%%%%%%%%%%%%%%%%%%%%%%%%%%%%%%%%%%%%%%%%%%%%%%%%%%%%%%%%%%%%%%%%%%%%%%%%%%%%%%%%%%%% 
\begin{figure*}[t]
\resizebox{3.8in}{!}{\includegraphics{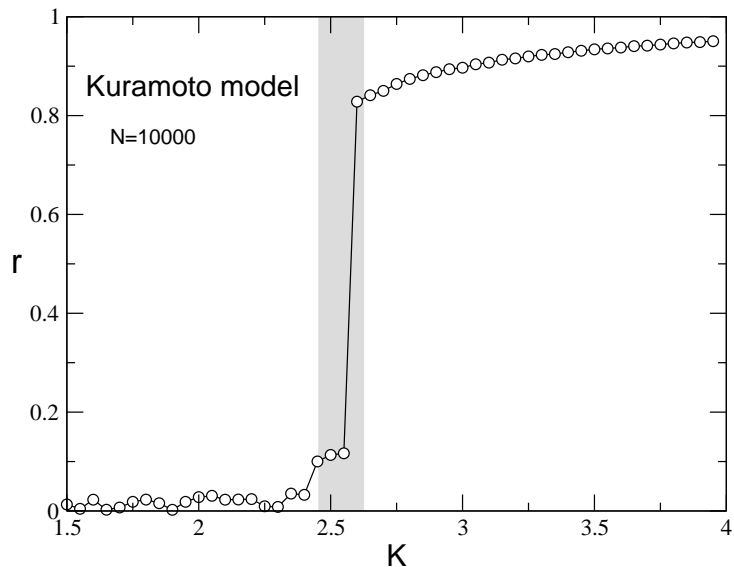}}
\caption{For the Kuramoto model we plot the asymptotic order parameter $r$ as a function 
of the coupling strenght $K$. We considered N=10000 and an average over 10 runs. The grey area  indicates the region where we have found the presence of metastable states. See text and next figure.}
\label{kura-1}
\end{figure*}
%%%%%%%%%%%%%%%%%%%%%%%%%%%%%%%%%%%%%%%%%%%%%%%%%%%%%%%%%%%%%%%%%%%%%%%%%%%%%%%%%%%%%%%%%%
%%%%%%%%%%%%%%%%%%%%%%%%%%%%%%%%%%%%%%%%%%%%%%%%%%%%%%%%%%%%%%%%%%%%%%%%%%%%%%%%%%%%%%%%% 
\begin{figure*}[t]
 \resizebox{4.15in}{!}{ \includegraphics{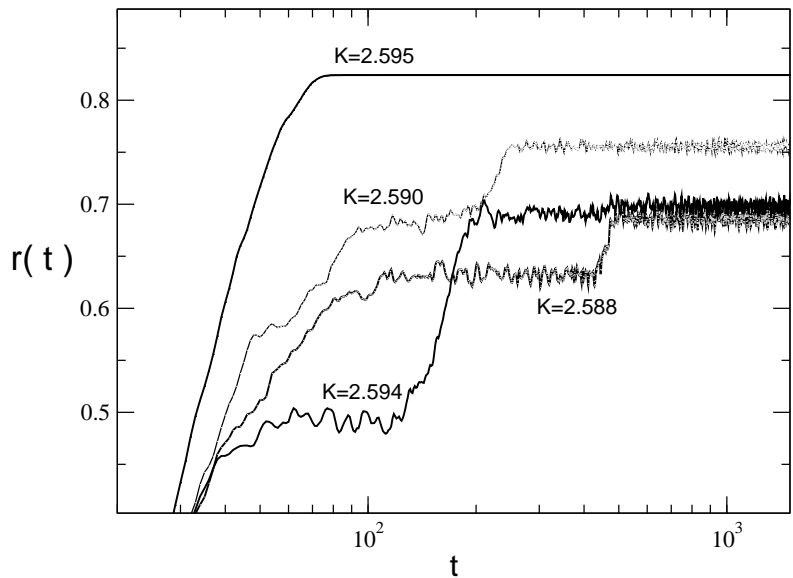}  }
\caption{For the Kuramoto model, we show the metastable states found for values of $K$ taken inside the grey area of the previous figure. The number of oscillator is N=10000 and an 
average over 10 runs  was considered.}
\label{kura-2}
\end{figure*}
%%%%%%%%%%%%%%%%%%%%%%%%%%%%%%%%%%%%%%%%%%%%%%%%%%%%%%%%%%%%%%%%%%%%%%%%%%%%%%%%%%%%%%%%%%

\section{Metastability in the Kuramoto model}

One of the the simplest 
and most successful models for synchronization is the famous 
Kuramoto model of coupled oscillators  \cite{kuramoto_model,strogatz}. 
It is simple enough to be analytically solvable, still retaining 
the basic principles to produce a rich variety of dynamical regimes
and synchronization patterns. 
The dynamics of the model is given by
\begin{equation}
    {\theta_i}' (t)  = \omega_i + \frac{K}{N} \sum_{j=1}^N
      \sin ( \theta_j  - \theta_i )  ~~~~~i=1,\dots,N
\label{kuramoto_eq1}
\end{equation}
where $\theta_i (t)$ is the phase (angle) of the $i$th oscillator at time $t$ ($-\pi<\theta_i(t)<\pi$),
while $\omega_i$ is its intrinsic frequency randomly drawn from
a symmetric, unimodal distribution $g(\omega)$ with a first moment
$\omega_0$ (typically a Gaussian distribution or a uniform one). These
natural frequencies $\omega_i$  are time-independent. 
The sum in the above equation runs over all the
oscillators so that the model  is an example of a globally coupled system. 
The parameter $K \geq 0$ measures the coupling strength: 
for small  values of $K$, each oscillator tends to run independently with its own
frequency,  while for large values of  $K$,  the coupling tends to synchronize (in phase and frequency) the oscillator with all the others.
Kuramoto showed that the model,
despite the difference in the natural frequencies of the
oscillators, exhibits a
spontaneous transition from incoherence to collective synchronization,
as the coupling strength is increased beyond  a certain critical threshold $K_c$ \cite{strogatz}.
\\
In Fig.2  we  plot  the asymptotic order parameter  
$r=\frac{1}{N}\sum_j^N e^{i \theta_j}$ versus the coupling $K$. In the simulations we used a uniform distribution for the natural frequencies and a  homogeneous initial conditions for the phases, i.e. $r(0)=0$. The figure shows a sharp transition from a homogenous non-synchronized state to a synchronized regime above $K_c\sim 2.6$. 
We considered $N=10000$ and  average over 10 runs.   Also in this case one can observe the emergence of metastable states just below the critical point.  The grey area indicates  the region  where this phenomenon has been found. 
In Fig. 3  we show the time evolution 
of the order parameter for different values of the coupling $K$ in the metastability region.
   Also in this case the lifetime of the QSS seems to depend on the size of the system in  close analogy with the metastable states found in the HMF model. 
 A detailed study of this behavior is in progress and will be reported elsewhere.   
 
   Despite the non hamiltonian character of eq.(3),  there is a formal link between the Kuramoto and the HMF model. 
Actually, both models can be considered as particular cases of the following general equation describing a generic pendulum driven by a constant torque $\Gamma$:
\begin{equation}
A {\theta}'' + B {\theta}'+ C sin(\theta) = \Gamma
\label{pendulum}
\end{equation}
where $B$ is a viscous damping constant while $A$ and $C$ are constants depending on mass, lenght and force.
In fact in the conservative case ($B= \Gamma = 0$), from eq.(\ref{pendulum}) we recover the equations of motion of HMF model rewritten using the order parameter $M$, with 
 the global angle $\Phi=0$ and considering each spin (pendulum) as moving in a mean-field potential determined by the istantaneous positions of all the other spins (pendula):
\begin{equation}
{\theta_i}''= -M sin(\theta_i)~~~~~~~~i=1,...,N.
\label{pendHMF}
\end{equation}
On the other hand, if we consider the overdamped limit (when $A$ is negligible with respect to the friction coefficient B), Eq.(\ref{pendulum}) reduces to the Kuramoto  equation, written also in this case using the order parameter $r$ with the the global phase equal to zero:
\begin{equation}
{\theta_i}'= \omega_i - K~r~sin(\theta_i)~~~~~~~i=1,...,N. 
\label{pendKura}
\end{equation}
In the latter case the natural frequencies $\omega_i$ play the role of the constant torque $\Gamma$.

These formal  similarities provide further support to the possibility of a common  explanation of metastability in terms of  a dynamical hindrance of synchronization.

\section{Conclusions}

Concluding this short review, we want to stress again the importance of the dynamics to understand  the origin of metastability for the HMF model  and the fundamental role played
by the initial conditions, 
which are able in some cases to produce a non homogeneous dynamics hardly tractable within a Vlasov approach. This has lead us to follow  several approaches which differ   from those used in refs. \cite{yama,bouchet,barre,rocha}.  Tsallis statistics, glassy dynamics and hindrance of synchronization  provide very interesting perspectives and promising  routes to be further explored in the future for 
a better understanding of the   dynamical origin of metastability
and complex behaviour in fully coupled systems.

\begin{acknowledgments}
We would like to thank S. Boccaletti, F. Bouchet,  P-H. Chavanis, T.Dauxois, A. Giansanti, V. Latora, D. Mukamel,   S. Ruffo and C. Tsallis for interesting  and stimulating discussions.
\end{acknowledgments}

\onecolumngrid


\begin{thebibliography}{99}


\bibitem{lhs1} T. Dauxois, V. Latora, A. Rapisarda, S. Ruffo and A. Torcini,  in {\it Dynamics   and     Thermodynamics   of Systems with Long Range Interactions}, T. Dauxois, S. Ruffo, E. Arimondo, M. Wilkens eds.,  Lecture Notes in Physics Vol. 602,   Springer   (2002) p. 458 and refs. therein.
 
 \bibitem{lrt} V. Latora, A. Rapisarda and C. Tsallis, Phys. Rev. E  64  (2001) 056134; V.Latora, A.Rapisarda and A. Robledo, Science  300 (2003) 249.
 
 
 \bibitem{cmt}A. Pluchino,  V. Latora,  A. Rapisarda, 
 Physica D 193 (2004) 315, Continuum Mechanics and Thermodynamics   16 (2004)  245 and  cond-mat/0507005.
 
 
\bibitem{monte} M.A. Montemurro, F.A. Tamarit and C. Anteneodo, Phys. Rev. E 67,  (2003) 031106 .  
 
\bibitem{yama} Y.Y. Yamaguchi, J. Barr\'e, F. Bouchet, T. Dauxois, S. Ruffo, Physica A 337  (2004) 36.

\bibitem{bouchet} F. Bouchet, T. Dauxois, Phys. Rev. E 72 (2005) 045103(R). 

\bibitem{barre} J. Barr\'e et al. cond-mat/0511070.
 
 \bibitem{rocha} T.M. Rocha Filho,  A. Figueiredo, M.A. Amato, cond-mat/0510056. 
 
 \bibitem{chava1}  P-H. Chavanis, J. Vatterville, F. Bouchet,   Eur. Phys. Jour.  B  46 (2005) 61.
 
 \bibitem{chava2}  P-H. Chavanis, cond-mat/0509767.
 
 \bibitem{morita} H. Morita and K. Kaneko, Europhys. Lett. 66 (2004) 198.
 
 \bibitem{gian} For a generalization of the HMF model see:  A. Campa, A. Giansanti,  D. Moroni, J. Phys. A: Math. Gen. 36 (2003) 6897 and refs. therein.
 
 \bibitem{hmfvetri}A. Pluchino,  V. Latora,  A. Rapisarda,  Phys. Rev. E 69 (2004) 056113,  cond-mat/0506665 and A. Pluchino and A. Rapisarda, Progr. Theor. Phys. (2005) in press, cond-mat/0509031 .
 
 \bibitem{epnhmf1} A. Rapisarda and A. Pluchino, Europhysics News,  36 (2005) 202  
 
 \bibitem{epnhmf2} F. Tamarit  and C. Anteneodo, Europhysics News,  36 (2005) 194 
 

\bibitem{tsa1}  Tsallis  J. Stat. Phys. 52 (1988) 479. See also  the special issue of Europhysics News, 36 (2005). 
 

\bibitem{kuramoto_model}
Y. Kuramoto, in {\it International Symposium on Mathematical Problems in
Theoretical Physics}, Vol.~39 of {\it Lecture Notes in Physics}, edited 
by H. Araki (Springer-Verlag, Berlin, 1975).

\bibitem{strogatz}
S.~H Strogatz,  Physica D, {\bf 143}  (2000) 1.

\end{thebibliography}
\end{document}